\documentclass[
showpacs,preprintnumbers,amsmath,amssymb]{revtex4}
\usepackage{amsmath}
\usepackage{graphicx}
\usepackage{subfig}
\usepackage{hyperref}
\usepackage{amssymb}
\usepackage[english]{babel}
\usepackage{epsfig}
\usepackage{wasysym}
\usepackage{graphicx}
\usepackage{indentfirst}
\usepackage{amsmath}
\usepackage{amsfonts}
\usepackage{amssymb}
\usepackage{enumerate}

\hypersetup{colorlinks,citecolor=green,filecolor=magenta,linkcolor=red,urlcolor=cyan,pdftex}

\newcommand{\be}{\begin{equation}}
\newcommand{\ee}{\end{equation}}
\newcommand{\bea}{\begin{eqnarray}}
\newcommand{\eea}{\end{eqnarray}}
\newcommand{\beaa}{\begin{eqnarray*}}
\newcommand{\eeaa}{\end{eqnarray*}}

\begin{document}


\title{Unimodular $f(G)$ gravity}

 \author{  M. J. S. Houndjo$^{(a,b)}$\footnote{e-mail:sthoundjo@yahoo.fr}  }

  
\affiliation{$^{a}$\, Facult\'e des Sciences et Techniques de Natitingou, BP 72, Natitingou, B\'enin \\
$^b$ \, Institut de Math\'{e}matiques et de Sciences Physiques, 01 BP 613,  Porto-Novo, B\'{e}nin} 

\begin{abstract}
In this paper we study a modified version of unimodular general relativity in the context of $f(G)$, $G$ denoting the Gauss-Bonnet invariant. We attach attention to Bianchi-type I and Friendmann-Robertson-Walker universes and  search for unimodular $f(G)$ models according to the de Sitter and power-law solutions. Assuming unimodular $f(G)$ gravity as perfect fluid and making use of the slow-roll parameters, inflationary model has been reconstructed in concordance with the Planck observational data. Moreover, we investigate the realization of the bounce and loop quantum cosmological ekpyrotic paradigms. Assuming suitable and appropriated scale factors, unimodular $f(G)$ models able to reproduce superbounce and ekpyrotic scenarios have been reconstructed.

\end{abstract}
\pacs{98.80.-k, 04.50.Kd, 98.80.Cq}

\maketitle 

\section{Introduction}
It is well known nowadays that interest in attached to gravity theories derived from Lagrangians extended beyond the General Relativity (GR). The first step of this is introducing the cosmological constant, originating from the vacuum expectation value of quantum field \cite{1deNOO,NOO}. However there is no intrinsic mechanism in theory that can dynamically induce the cosmological constant \cite{23deBOS,BOS}. On the other hand, unimodular gravity \cite{2deNOO}-\cite{26deNOO} is an interesting gravitational theory that can be considered as a specific case of the GR, and in which the cosmological constant appears as the trace-free part of the gravitational field equations, fixing the determinant of the metric tensor as a number or a function\cite{NOO,BOS}. In other-word it would be more appropriate to state that unimodular gravity, in a way, can offer a proposal to solve cosmological problem. It is important to mention that in the context of unimodular gravity the problem of late-time acceleration of the universe can be developed,  where the metric can be decomposed in the unimodular metric part and a scalar field part\cite{20deNOO}-\cite{22deNOO}. Instead of considering GR, different kinds of modified gravity based on the curvature scalar have been performed in the recent years, as $f(R)$ \cite{2deHRMM}-\cite{5deHRMM} where $R$ is the curvature scalar, the $f(R,T)$, $T$ being the trace of the energy-momentum tensor \cite{houndjo1}-\cite{houndjo2}, and $f(G)$, $G$ denoting the Gauss-Bonnet (GB) invariant \cite{6deHRMM}-\cite{HRMM}.\par
In this paper we focus our attention to the $f(G)$ and the purpose here is to realize some cosmological evolutions with specific and realistic Hubble rate, and investigate which unimodular $f(G)$ model can yield such cosmological evolutions of the universe. We fundamentally adopt two kinds of metric in this paper; the Bianchi type-I (BI) metric and the Friedman-Robertson-Walker (FRW). It is obvious that for both metrics there is no compatibility with the unimodular constraint, and so, we fix suitably the metrics in order to satisfy this constraint. With the BI metric we search for $f(G)$ models able to reproduce inflation through the de Sitter and power-law form of the scale factor. Through the use of the FRW metric, we still search for $f(G)$ models that can state the inflation. As we are leading with inflationary models we find important to formulate the related-observables, as the spectral index, the tensor-to-tensor ratio and the running of the spectral index. By considering $f(G)$ model coming from power-law solutions of the scale factor, with non specified integration constants in the perfect fluid description, we determine the slow-roll indice and confront them with the observational recent Planck 
\cite{78deNOO,79deNOO} in order to suitably calculate the constants, and then obtaining the effective unimodular $f(G)$ model able to realize the inflationary epoch of our universe. \par
An alternative to the standard description of the acceleration of inflation is provided by bouncing cosmologies \cite{9deOOS,10deOOS}, in which the appearance of the initial singularity is prevented. This feature can be represented by the use of scalar fields \cite{9deOOS}-\cite{22deOOS}, by the models of modified gravity \cite{23deOOS}-\cite{35deOOS}, and also by the use of the Loop Quantum Cosmology (LQC) \cite{36deOOS}-\cite{44deOOS}. In the general case, the realization of the bounce is related to the coupling of matter fields with an equation of state in order to violate the null energy condition \cite{9deOOS}-\cite{12deOOS}. Bouncing cosmologies have been developed in several works \cite{45deOOS}-\cite{47deOOS,OOS}, but essentially, it has been shown that bouncing cosmologies may present primordial instabilities \cite{13deOOS,14deOOS}, that can be solved by the ekpyrotic scenario. In this paper we also attach attention to the superbounce and the LQC ekpyrotic scenarios in the context of unimodular $f(G)$ gravity by reconstructing the suitable model characteristic of each scenario. This kind of work has been performed in \cite{OOS}, but not in the unimodular context. For other reviews about bouncing cosmologies see \cite{72deOOS,73deOOS} for $f(R)$ theory, \cite{74deOOS}-\cite{83deOOS,86deOOS,87deOOS} for $f(G)$ theory and \cite{88deOOS}-\cite{113deOOS} for $f(T)$ theory, where $T$ denotes the torsion scalar.\par
The manuscript is organized as follows: in Sec.\ref{sec2} we present the general description of unimodular gravity. The Sec.\ref{sec3} is devoted to the formalism of $f(G)$ and unimodular equations of motion. The reconstruction of $f(G)$ gravity in inhomogeneous universe is performed in Sec.\ref{sec4}, for both de Sitter solutions in Subsec.\ref{subsec41} and power-law solutions in Subsec.\ref{subsec42}. In the FRW metric context, the reconstruction of unimodular $f(G)$ models have been developed within the de Sitter solutions in Subsec.\ref{subsec51} and power-law solutions in Subsec.\ref{subsec52}, and this later has been achieved according to Planck results in Sec.\ref{sec6} finding the integration constants accordingly. The ekpyrotyc scenario reconstruction  and superbounce reconstruction from unimodular $f(G)$ gravity have been performed in the Sec.\ref{sec7}. We present our conclusion in Sec.\ref{sec8}.

\section{General description of unimodular gravity}\label{sec2}
In this section we point out the generalization of the GR gravity formalism in order to provide the unimodular $f(G)$ gravity formalism. The unimodular gravity approach is based on the assumption that the determinant of the metric tensor is fixed, expressed by the relation $g_{\mu\nu}\delta g^{\mu\nu}=0$. Throughout this paper the components of the metric is chosen such a way so that \cite{NOO}
\begin{eqnarray}\label{cond1} 
\sqrt{-g}=1
\end{eqnarray}
Let us start our study by using the Bianchi type-I metric, from which the usual and practical FRW metric can be recovered, as 
\begin{eqnarray}\label{metric1}
ds^2=dt^2-A(t)^2dx^2-B(t)^2dy^2-C(t)^2dz^2
\end{eqnarray}
It is easy to check that the unimodular constraint, expressed by the Eq. (\ref{cond1}) is not satisfied by the Bianchi-type I metric (\ref{metric1}). In order to reach this condition we need to redefine the cosmic time coordinate as follows
\begin{eqnarray}
d\tau = A(t)B(t)C(t)dt,\label{ttau}
\end{eqnarray} 
such a way that the metric (\ref{metric1}) becomes
\begin{eqnarray}\label{metric2}
ds^2= \left[A(t(\tau))B(t(\tau))C(t(\tau))\right]^{-2}d\tau^2-A(t(\tau))^2dx^2-B(t(\tau))^2dy^2-C(t(\tau))^2dz^2.
\end{eqnarray}
This metric obviously satisfies the constraint (\ref{cond1}) and we shall refer to this later as the unimodular Bianchi-type I metric from which the FRW metric can be recovered. 

\section{Formalism of $f(G)$ gravity and unimodular equations of motion}\label{sec3}

Let us introduce the general $R+f(G)$ action as follows
\begin{eqnarray}\label{fgaction}
S=\frac{1}{2\kappa}\int d^4x\sqrt{-g}\left[R+f(G)\right]+S_m ,
\end{eqnarray}
where $R$ represents the Ricci scalar, and the modification function $f(G)$ corresponds to a generic globally differentiable Gauss-Bonnet topological invariant $G$ function. The matter action $S_m$ is the one which induces the energy momentum tensor $T_{\mu\nu}$.
We focus our attention on the metric formalism where the variation of the action (\ref{fgaction}) with respect to the metric tensor yields within the minimum principle the following general equation of motion \cite{6deHRMM}
\begin{eqnarray}
R_{\mu\nu}-\frac{1}{2}Rg_{\mu\nu}+8\Bigg[R_{\mu\rho\nu\sigma}+R_{\rho\nu}g_{\sigma\mu}-R_{\rho\sigma}g_{\nu\mu}-R_{\mu\nu}g_{\sigma\rho}+R_{\mu\sigma}g_{\nu\rho}\nonumber\\+\frac{1}{2}R\left(g_{\mu\nu}g_{\sigma\rho}-g_{\mu\sigma}g_{\nu\rho}\right)\Bigg]\nabla^{\rho}\nabla^{\sigma}f_G+\left(Gf_G-f\right)g_{\mu\nu}=\kappa T_{\mu\nu}.
\end{eqnarray}
Here $f_G=\frac{df(G)}{dG}$, and the Gauss-Bonnet term is defined by $G=R^2-R_{\mu\nu}R^{\mu\nu}+R_{\mu\nu\lambda\sigma}R^{\mu\nu\lambda\sigma}$, $R_{\mu\nu}$ and $R_{\mu\nu\lambda\sigma}$ being the Ricci tensor and Riemann tensor, respectively. We also point out some adopted definitions as follows: the signature of the Riemannian metric is $(+---)$, $\nabla_{\mu}V_\nu=\partial\mu V_\nu-\Gamma^{\lambda}_{\mu\nu}V_\lambda$ and $R^{\sigma}_{\mu\nu\rho}=\partial_{\nu}\Gamma^{\sigma}_{\mu\rho}-\partial_{\rho}\Gamma^{\sigma}_{\mu\nu}+\Gamma^{\omega}_{\mu\rho}\Gamma^{\sigma}_{\omega\nu}$  for the covariant derivative of a covariant vector and the Riemann tensor, respectively. Making use of the metric (\ref{metric2}) the equations of field read
\begin{eqnarray}
-24A^3A'B^3B'C^3C'f'_G+\Bigg(8A^3A'B^3B'C^3C''+24A^3A'B^3B'C^2C'^2+8A^3A'B^3B''C^3C'
\nonumber\\
+24A^3A'B^2B'^2C^3C'+8A^3A''B^3B'C^3C'+24A^2A'^2B^3B'C^3C'\Bigg)f_G\nonumber\\
+A^2BB'CC'+AA'B^2CC'+AA'BB'C^2-f=8\pi\rho
\end{eqnarray}
\begin{eqnarray}
8A^4B^3B'C^3C'f_G''+\Bigg(8A^4B^3B'C^3C''+24A^4B^3B'C^2C'^2+8A^4B^3B''C^3C'+24A^4B^2B'^2C^3C'\nonumber\\
+24A^3A'B^3B'C^3C'\Bigg)f_G'+\Bigg(-8A^3A'B^3B'C^3C''-24A^3A'B^3B'C^2C'^2-8A^3A'B^3B''C^3C\nonumber\\
'-24A^3A'B^2B'^2C^3C'-8A^3A''B^3B'C^3C'-24A^2A'^2B^3B'C^3C'\Bigg)f_G-A^2B^2CC''-A^2B^2C'^2\nonumber\\
-3A^2BB'CC'-AA'B^2CC'-A^2BB''C^2-A^2B'^2C^2-AA'BB'C^2+f=8\pi p_r
\end{eqnarray}
\begin{eqnarray}
8A^3A'B^4C^3C'f_G''+\Bigg(8A^3A'B^4C^3C''+24A^3A'B^4C^2C'^2+24A^3A'B^3B'C^3C'+8A^3A''B^4C^3C'\nonumber\\
24A^2A'^2B^4C^3C'\Bigg)f_G'+\Bigg(-A^3A'B^3B'C^3C''-24A^3A'B^3B'C^2C'^2-8A^3A'B^3B''C^3C'-24A^3A'B^2B'^2C^3C'\nonumber\\
-8A^3A''B^3B'C^3C'-24A^2A'^2B^3B'C^3C'\Bigg)f_G-A^2B^2CC''-A^2B^2C'^2-A^2BB'CC'\nonumber\\-3AA'B^2CC'-AA'BB'C^2-AA''B^2C^2-A'^2B^2C^2+f=8\pi p_t
\end{eqnarray}
\begin{eqnarray}
8A^3A'B^3B'C^4f_G''+\Bigg(24A^3A'B^3B'C^3C'+8A^3A'B^3B''C^4+24A^3A'B^2B'^2C^4+8A^3A''B^3B'C^4\nonumber\\
+24A^2A'^2B^3B'C^4\Bigg)f_G'+\Bigg(-8A^3A'B^3B'C^3C''-24A^3A'B^3B'C^2C'^2-8A^3A'B^3B''C^3C'-24A^3A'B^2B'^2C^3C'\nonumber\\
-8A^3A''B^3B'C^3C-24A^2A'^2B^3B'C^3C'\Bigg)f_G-A^2BB'CC'-AA'B^2CC'-A^2BB''C^2\nonumber\\
-A^2B'^2C^2-3AA'BB'C^2-AA''B^2C^2-A'^2B^2C^2+f=8\pi p_t,
\end{eqnarray}
where we undertook the matter content as an anisotropic fluid described by the following energy-momentum tensor
\begin{eqnarray}\label{emt}
T_{\mu\nu}=\left(\rho+p_t\right)u_\mu u_\nu-p_tg_{\mu\nu}+\left(p_r-p_t\right)v_\mu v_\mu,
\end{eqnarray}
with $\rho$ being the energy density, $u_\mu$ the 4-velocity and $v_\mu$ the space-like vector in the radial direction; $p_r$ (the radial pressure) is the pressure in the direction of $v_\mu$ and $p_t$ (the tangential pressure) the pressure in the direction orthogonal to $v_\mu$. Due to the fact of considering an anisotropic spherically symmetric matter one has $p_r\neq p_t$ (their equality corresponds to an isotropic spherically symmetric matter).
\section{Reconstructing $f(G)$ gravity in inhomogeneous universe}\label{sec4}
In this section we search for the $f(G)$ action through the reconstruction scheme  for some particular solution of the class of the metric explored in the previous section. We will just consider the de Sitter and power-law solutions. In a general case, using (\ref{metric2}) and (\ref{emt}) the conservation equation for the energy-momentum tensor can easily be obtained by
\begin{eqnarray}
\dot{\rho}+\left(H_x+H_y+H_z\right)\rho+H_xp_r+\left(H_y+H_z\right)p_t=0
\end{eqnarray}
where the following definitions have been made $H_x=\frac{\dot A}{A}$, $H_y=\frac{\dot B}{B}$ and $H_z=\frac{\dot C}{C}$
\subsection{Searching for de Sitter solutions}\label{subsec41}
The de Sitter solutions are ones of the well known in the context of the cosmology due to the fact that they may approximately describe the early and  current epochs of the universe where its expansion is accelerated. Here, for the three directions the scale factors present an exponential expansion yielding constant Hubble parameters in each direction. Then we assume the scale factors as follows
\begin{eqnarray} \label{desitterscale}
A=A_0e^{H_{x0}t},\quad\quad B=B_0e^{H_{y0}t}, \quad\quad C=C_0e^{H_{z0}t},
\end{eqnarray}
where , $H_{x0}$, $H_{y0}$ and $H_{z0}$ are positive constants expressing the instantaneous  rates of the expansion at $t=0$. By simple derivations one can see that the rates of the expansion for each directions read
\begin{eqnarray}
H_x=\frac{\dot A}{A}=H_{x0}, \quad\quad H_y=\frac{\dot B}{B}=H_{y0}, \quad\quad H_x=\frac{\dot C}{C}=H_{z0}\;.
\end{eqnarray} 
Then, in the context of the cosmic time $t$ the Gauss Bonnet invariant reads
\begin{eqnarray}
G&=&\frac{8}{ABC}\left(  \dot A \dot B \ddot C +\dot A \ddot B \dot C +\ddot A \dot B \dot C   \right)\\
&=& 8H_{x0}H_{y0}H_{z0}\left(H_{x0}+H_{y0}+H_{z0}\right)
\end{eqnarray}
 and the field equations become
\begin{eqnarray}
8H_{x0}H_{y0}H_{z0}\left(H_{x0}+H_{y0}+H_{z0}\right)f_G+H_{x0}H_{y0}+H_{x0}H_{z0}+H_{y0}H_{z0}-f=8\pi \rho\\
-8H_{x0}H_{y0}H_{z0}\left(H_{x0}+H_{y0}+H_{z0}\right)f_G-H_{z0}^2-H_{y0}H_{z0}-H_{y0}^2+f= 8\pi p_r\\
-8H_{x0}H_{y0}H_{z0}\left(H_{x0}+H_{y0}+H_{z0}\right)f_G-H_{z0}^2-H_{x0}H_{z0}-H_{x0}^2+f= 8\pi p_t\\
-8H_{x0}H_{y0}H_{z0}\left(H_{x0}+H_{y0}+H_{z0}\right)f_G-H_{y0}^2-H_{x0}H_{y0}-H_{x0}^2+f= 8\pi p_t
\end{eqnarray}
Since we are leading with de Sitter solutions, by assuming  $p_r=p_t=p$, one has $p=-\rho$, and the combination of any two of the field equations yields
\begin{eqnarray}\label{eqdiff1}
G\frac{df(G)}{dG}-f(G)+K=0\,,
\end{eqnarray} 
where $K$ is a constant depending on $H_{x0}$, $H_{y0}$ and $H_{z0}$. The general solution of the equation (\ref{eqdiff1})  reads
\begin{eqnarray}
f(G)=\alpha G +K,\label{fGdesitter1}
\end{eqnarray}
where $K$ is an integration constant. Making use of (\ref{ttau}) and (\ref{desitterscale}) the expressions of the scale factors can written as 
\begin{eqnarray}
A[t(\tau)]=\left(\frac{H_{x0}+H_{y0}+H_{z0}}{B_0C_0}\tau\right)^{\frac{H_{x0}}{H_{x0}+H_{y0}+H_{z0}}}\\
B[t(\tau)]=\left(\frac{H_{x0}+H_{y0}+H_{z0}}{A_0C_0}\tau\right)^{\frac{H_{y0}}{H_{x0}+H_{y0}+H_{z0}}}\\
A[t(\tau)]=\left(\frac{H_{x0}+H_{y0}+H_{z0}}{A_0B_0}\tau\right)^{\frac{H_{z0}}{H_{x0}+H_{y0}+H_{z0}}}\;,
\end{eqnarray} 
such that the metric (\ref{metric2}) in the unimodular context becomes
\begin{eqnarray}
ds^2=Q\tau^{-2}d\tau^2-Q_1\tau^{q_1}dx^2-Q_2\tau^{q_2}dy^2-Q_3\tau^{q_3}dz^2
\end{eqnarray}
with 
\begin{eqnarray}
Q=\left(H_{x0}+H_{y0}+H_{z0}\right)^{-2}\left(B_0C_0\right)^{q_1}\left(A_0C_0\right)^{q_2}\left(A_0B_0\right)^{q_3}\\
Q_1=A_0^2\left(\frac{H_{x0}+H_{y0}+H_{z0}}{A_0B_0C_0}\right)^{q_1}\;,\quad q_1=\frac{2H_{x0}}{H_{x0}+H_{y0}+H_{z0}}\\
Q_2=B_0^2\left(\frac{H_{x0}+H_{y0}+H_{z0}}{A_0B_0C_0}\right)^{q_2}\;,\quad q_2=\frac{2H_{y0}}{H_{x0}+H_{y0}+H_{z0}}\\
Q_3=C_0^2\left(\frac{H_{x0}+H_{y0}+H_{z0}}{A_0B_0C_0}\right)^{q_3}\;,\quad q_3=\frac{2H_{z0}}{H_{x0}+H_{y0}+H_{z0}}
\end{eqnarray}
\subsection{Power-law solutions}\label{subsec42}
In this subsection we devote our attention to the cosmological evolution described by a power-law functions of cosmic time in each direction of the space. We assume the following expressions for the scale factors in terms of cosmic time as
\begin{eqnarray}
A(t)=A_0t^a\;,\quad B(t)=B_0t^b\;,\quad C(t)=C_0t^c\;,
\end{eqnarray}
where ${a,b,c}$ and ${A_0, B_0, C_0}$ are constants that should be determined from the initial conditions. In this case the rate of the expansion for each direction is given by
\begin{eqnarray}
H_x=\frac{a}{t}\;,\quad\quad H_y=\frac{b}{t}\;,\quad\quad H_z=\frac{c}{t}
\end{eqnarray}
and the expression of Gauss-Bonnet invariant in the context of cosmic time reads
\begin{eqnarray}\label{BGPL}
G=\frac{8\,a\,b\,{c}^{2}}{{t}^{4}}+\frac{8\,a\,{b}^{2}\,c}{{t}^{4}}+\frac{8\,{a}^{2}\,b\,c}{{t}^{4}}-\frac{24\,a\,b\,c}{{t}^{4}}.
\end{eqnarray}
Then the fields equations take the following forms
\begin{eqnarray}
Gf_G+\frac{b\,c}{{t}^{2}}+\frac{a\,c}{{t}^{2}}+\frac{a\,b}{{t}^{2}}-f=8\,\pi \,\rho\\
-Gf_G-\frac{\left( c-1\right) \,c}{{t}^{2}}-\frac{b\,c}{{t}^{2}}-\frac{\left( b-1\right) \,b}{{t}^{2}}+f=8\,\pi\,p_r\\
-Gf_G-\frac{\left( c-1\right) \,c}{{t}^{2}}-\frac{a\,c}{{t}^{2}}-\frac{\left( a-1\right) \,a}{{t}^{2}}+f= 8\,\pi\,p_t\\
-Gf_G-\frac{\left( b-1\right) \,b}{{t}^{2}}-\frac{a\,b}{{t}^{2}}-\frac{\left( a-1\right) \,a}{{t}^{2}}+f=8\,\pi\,p_t
\end{eqnarray}
Here we also assume for simplicity that $p_r=p_t=p=\omega\rho$, $\omega$ being the parameter of the equation of state. Considering the first two field equations one gets
\begin{eqnarray}
G\left(1+8\pi \omega\right)f_G-\left(1+8\pi\omega\right)f+\left({K}_2+8\pi\omega{K}_1\right)\frac{1}{t^2}=0 
\end{eqnarray} 
with $K_1={b\,c}+{a\,c}+{a\,b}$ and $K_2=-{\left( c-1\right) \,c}-{b\,c}-{\left( b-1\right) \,b}$. By making use of (\ref{BGPL})  the previous equations takes the following form
\begin{eqnarray}
Gf_G-f+\mathcal{K}G^{1/2}=0
\end{eqnarray}
where
\begin{eqnarray}
\mathcal{K}=\frac{K_2+8\pi\omega K_1}{\left(1+8\pi\omega\right)\left( {8\,a\,b\,{c}^{2}}+{8\,a\,{b}^{2}\,c}+{8\,{a}^{2}\,b\,c}-{24\,a\,b\,c}\right)^{1/2}  }.
\end{eqnarray}
The general solution of the previous equation reads
\begin{eqnarray}
f(G)=2\mathcal{K}\sqrt{G}+\beta G
\end{eqnarray}
with $\beta$ a positive constant to be determined using the cosmological data. In this case the metric (\ref{metric2}) in the unimodular context becomes
\begin{eqnarray}
ds^2=\mathcal{Q}\tau^{r}d\tau^2-\mathcal{Q}_1\tau^{r_1}dx^2-\mathcal{Q}_2\tau^{r_2}dy^2-\mathcal{Q}_3\tau^{r_3}dz^2
\end{eqnarray}
where 
\begin{eqnarray}
\mathcal{Q}=\frac{1}{(A_0B_0C_0)^2}\left(\frac{a+B+C+1}{A_0B_0C_0}\right)^r\:,\quad r=-2\frac{a+b+c}{a+b+c+1}\:,\quad \mathcal{Q}_1=A_0^2\left(\frac{a+B+C+1}{A_0B_0C_0}\right)^{r_1}\\
r_1=\frac{2a}{a+b+c+1}\;,\quad  \mathcal{Q}_2=B_0^2\left(\frac{a+B+C+1}{A_0B_0C_0}\right)^{r_2}\;,\quad r_2=\frac{2b}{a+b+c+1}\\
\mathcal{Q}_3=C_0^2\left(\frac{a+B+C+1}{A_0B_0C_0}\right)^{r_3}\;,\quad r_3=\frac{2c}{a+b+c+1} 
\end{eqnarray}

\section{ Lagrange multiplier formulation and reconstruction of unimodular $f(G)$ gravity}\label{sec5}

In this section we will present the Lagrange multiplier formulation of $f(G)$ gravity, according to the method performed in context of $f(R)$ in \cite{26deBOS}. This procedure is just used to ensure that the unimodular condition is satisfied. To do so, we introduce the Lagrangian multiplier $\lambda$ and the unimodular $f(G)$ gravity with matter can be expressed as
\begin{eqnarray}
S_{\lambda}= \int d^4x\left\lbrace\sqrt{-g}\left[\frac{R+f(G)}{2\kappa^2}-\lambda\right]+\lambda\right\rbrace 
\end{eqnarray}
From now we will assume the FRW metric and consider the matter energy-momentum tensor as corresponding to a perfect fluid with the energy density and pressure $\rho$ and $p$, respectively. Thereby, the GB invariant expressions  in terms of $t$ and $\tau$ read
\begin{eqnarray}\label{GBI}
G&=&24\frac{\dot{A}^2\ddot{A}}{A^3}\\&=&24A^9A'^2A''+72A^8A'^4=24A^{12}\left(\mathcal{H}^2\mathcal{H}'+4\mathcal{H}^4\right)\label{GBItau}
\end{eqnarray}
where the parameter$\mathcal{H}\equiv\frac{1}{A(\tau)}\frac{dA(\tau)}{d\tau}$ is the relative Hubble parameter in this context. Hence the field equations become
\begin{eqnarray}
-24A^{12}\mathcal{H}^3f'_G+24A^{12}\left(\mathcal{H}^2\mathcal{H}'+4\mathcal{H}^4\right)f_G+3A^6\mathcal{H}^2-\left(f-\lambda\right)-\rho=0\label{frw1}\\
8A^{12}\mathcal{H}^2f''_G+\left(16A^{12}\mathcal{H}\mathcal{H}'+88A^{12}\mathcal{H}^3\right)f'_G-24A^{12}\left(\mathcal{H}^2\mathcal{H}'+4\mathcal{H}^4\right)f_G-A^6\left(2\mathcal{H}'+9\mathcal{H}^2\right)+\left(f-\lambda\right)-p=0.\label{frw2}
\end{eqnarray}
By directly summing the equations (\ref{frw1}) and (\ref{frw2}), the term $\left(f-\lambda\right)$  is eliminated giving rise to the following equations
\begin{eqnarray}
8A^{12}\mathcal{H}^2f''_G+16A^{12}\left(\mathcal{H}\mathcal{H}'+4\mathcal{H}^3\right)f'_G-2A^{6}\left(\mathcal{H}'+3\mathcal{H}^2\right)-\rho-p=0.\label{tobesolved}
\end{eqnarray}
In order to perform a consistent analysis, it is important to make use of the continuity equation for the matter, namely
\begin{eqnarray}
\rho'+3\mathcal{H}\left(\rho+p\right)=0
\end{eqnarray}
Let us adopt the equation of state $p=w\rho$. Thus solving the previous equation, one gets
\begin{eqnarray}
\rho(\tau)=\rho_0\exp\left\{-3\left(1+\omega\right)\int_0^\tau\mathcal{H}(\zeta)d\zeta\right\}=\rho_0A(\tau)^{-3(1+w)}
\end{eqnarray}
Now, in order to obtain cosmological $f(G)$ models in the unimodular context, one just need an suitable expression of the scale depending on $\tau$, solving the equation (\ref{tobesolved}).
\subsection{Reconstruction of unimodular $f(G)$ model describing de Sitter Universe}\label{subsec51}
In this subsection we consider the FRW metric that describes a de Sitter expanding universe with the scale factor reading
\begin{eqnarray}
A(t)=e^{H_0t},
\end{eqnarray}
where $H_0$ is an arbitrary positive constant. By using the relation (\ref{ttau}), setting $B=C=A$, one can extract the $t$ in terms of $\tau$ as
\begin{eqnarray}
t=\frac{1}{3H_0}\ln{\left(3H_0\tau\right)},
\end{eqnarray}
such that the scale factor in terms of the $\tau$ reads
\begin{eqnarray}
A(t(\tau))=\left(3H_0\tau\right)^{1/3}
\end{eqnarray}
In this case, leading with de Sitter universe, $\omega=-1$ and the differential equation (\ref{tobesolved})  becomes
\begin{eqnarray}
3\tau\frac{d^2f_G(\tau)}{d\tau^2}+2\frac{df_G(\tau)}{d\tau}=0,
\end{eqnarray}
whose general solutions is
\begin{eqnarray}
f_G(\tau)=3\mathcal{C}_1\tau^{1/3}+\mathcal{C}_2
\end{eqnarray}
where $\mathcal{C}_1$ and $\mathcal{C}_2$ are integration constants. On the other hand, one can expression the GB invariant (\ref{GBItau}) in terms of $\tau$ as
\begin{eqnarray}
G(\tau)=24H_0
\end{eqnarray}
which obviously is a constant. The task to be done now is to inverse the function $G(\tau)$ in order to obtain $\tau=\tau(G)$, but since $G(\tau)=Const$ it is commode to fix $\mathcal{C}_1=0$, such that $f_G=df(G)/dG=\mathcal{C}_2$, whose general solution takes the following form
\begin{eqnarray}
f(G)=\mathcal{C}_2G+\mathcal{C}_3
\end{eqnarray}
This result is consistent with the one obtained in (\ref{fGdesitter1}), confusing $(\alpha,K)$ with $(\mathcal{C}_1,\mathcal{C}_3)$, characteristic of the de Sitter Universe.
\subsection{Reconstruction of unimodular $f(G)$ model according to power-law solution}\label{subsec52}
In this case we assume the scale factor in terms of the cosmic time as
\begin{eqnarray}
A(t)=A_0\left(\frac{t}{t_0}\right)^q\label{scaletq}
\end{eqnarray}
where $t_0$ and $A_0$ are the initial time  and the initial value of the scale factor, respectively. According to the relation $d\tau=A(t)^3dt$ one gets
\begin{eqnarray}
A(t(\tau))=\left(\frac{\tau}{\tau_0}\right)^{\sigma},\quad\quad \sigma=\frac{q}{1+3q} \quad\quad \tau_0=\frac{t_0}{A_0^{1/q}(1+3q)}\label{scaletausigma}
\end{eqnarray}
It is easy to see that for $\sigma=1/3$, the de Sitter universe is recovered, while the power-law inflation is obtained for $q>1$ $(1/4<\sigma<1/3)$. The accelerated expansion is realized for $0<q\leq1$ $(0<\sigma\leq 1/4)$, where inflation is possible.\par
The GB invariant in this case is
\begin{eqnarray}
G=24\sigma^3\left(\frac{4\sigma-1}{\tau^4}\right)\left(\frac{\tau}{\tau_0}\right)^{12\sigma}\label{FRWGBI}
\end{eqnarray}
Thus the differential equation (\ref{tobesolved}) becomes
\begin{eqnarray}
J_{01}f''_G+J_{02}f'_G+(J_{03}+J_{04})=0.\label{tobesolved2}
\end{eqnarray}
with
\begin{eqnarray}
J_{01}(\tau)=\frac{8\sigma^2}{\tau^2}\left(\frac{\tau}{\tau_0}\right)^{12\sigma}\;,\quad
 J_{02}(\tau)=\frac{16\sigma^2(4\sigma-1)}{\tau^3}\left(\frac{\tau}{\tau_0}\right)^{12\sigma}\nonumber\\
 J_{03}(\tau)=-\frac{2\sigma(3\sigma-1)}{\tau^2}\left(\frac{\tau}{\tau_0}\right)^{6\sigma}\;,\quad 
J_{04}(\tau)=-(1+w)\rho_0\left(\frac{\tau}{\tau_0}\right)^{-3\sigma(1+w)}
\end{eqnarray}
The previous equation can be rewritten by dividing whole equation by $J{01}\neq 0$, as
\begin{eqnarray}
f''_G+J_{1}f'_G+J_2=0.\label{tobesolved3}
\end{eqnarray}
where
\begin{eqnarray}
J_1(\tau)=\frac{1-3\sigma}{\tau}\;,\quad\quad  J_2(\tau)=\frac{1-3\sigma}{4\sigma}\left(\frac{\tau}{\tau_0}\right)^{-6\sigma}+\frac{\rho_0(1+w)\tau^2}{8\sigma^2}\left(\frac{\tau}{\tau_0}\right)^{-3\sigma(5+w)}
\end{eqnarray}
whose general solution reads
\begin{eqnarray}
f_G(\tau)=\frac{\tau^2\left(\frac{\tau}{\tau_0}\right)^{-6\sigma}}{8\sigma(1-2\sigma)}+
\frac{\rho_0(1+w)\tau^4\left(\frac{\tau}{\tau_0}\right)^{-3\sigma(5+w)}}{8\sigma^2\left[ 4-3(5+w)\sigma \right]\left[-1+(7+3w)\sigma  \right]}+
\mathcal{K}_1\frac{\tau^{3-8\sigma}}{3-8\sigma}+\mathcal{K}_2,\label{fGdetau}
\end{eqnarray}
$\mathcal{K}_1$ and $\mathcal{K}_2$, being integration constants. From the relation (\ref{FRWGBI}), one can express $\tau$ and $\frac{\tau}{\tau_0}$ in terms of $G$ as
\begin{eqnarray}
\tau=\chi_1 G^{\frac{1}{4(3\sigma-1)}}\;,\frac{\tau}{\tau_0}=\chi_2 G^{\frac{1}{4(3\sigma-1)}}\quad \chi_1=\left[\frac{\tau_0^{12\sigma}}{24\sigma^3(4\sigma-1)}\right]^{\frac{1}{4(3\sigma-1)}}\;,\quad \chi_2=\left[\frac{\tau_0^{4}}{24\sigma^3(4\sigma-1)}\right]^{\frac{1}{4(3\sigma-1)}} .
\end{eqnarray}
Hence, the expression (\ref{fGdetau}) can now be written in terms of $G$ as
\begin{eqnarray}
f_G(G)=\frac{\chi_1^2\chi_2^{-6\sigma}}{8\sigma(1-2\sigma)}G^{-\frac{1}{2}}+
\frac{\rho_0(1+w)\chi_1^4\chi_2^{-3\sigma(5+w)}}{8\sigma^2\left[ 4-3(5+w)\sigma \right]\left[-1+(7+3w)\sigma  \right]}G^{\frac{4-3\sigma(5+w)}{4(3\sigma-1)}}+
\frac{\mathcal{K}_1\chi_1^{3-8\sigma}}{3-8\sigma}G^{\frac{3-8\sigma}{4(3\sigma-1)}}+\mathcal{K}_2.\label{fGdeG}
\end{eqnarray}
By integrating the expression in the right hand side of the (\ref{fGdeG}) with respect to $G$, one gets
\begin{eqnarray}
f(G)=\frac{\chi_1^2\chi_2^{-6\sigma}G^{\frac{1}{2}}}{4\sigma(1-2\sigma)}+
\frac{\rho_0(1-3\sigma)\chi_1^4\chi_2^{-3\sigma(5+w)}G^{\frac{3\sigma(1+w)}{4(1-3\sigma)}}}{6\sigma^3\left[ 4-3(5+w)\sigma \right]\left[-1+(7+3w)\sigma  \right]}+
\frac{4\mathcal{K}_1(3\sigma-1)\chi_1^{3-8\sigma}G^{\frac{4\sigma-1}{4(3\sigma-1)}}}{(4\sigma-1)(3-8\sigma)}+\mathcal{K}_2G+\mathcal{K}_3.\label{fdeG}
\end{eqnarray}

\section{Unimodular Inflationary Cosmological $f(G)$ model}\label{sec6}
In the previous section we performed the reconstruction of the specific $f(G)$ form that generates scale-factor evolution. The obtained model will be explored in the present section to investigate inflation realization in unimodular $f(G)$ gravity. Moreover, we will focus our attention on investigation of inflationary observables such as scalar and tensor spectral indices, the tensor-to-tensor ratio and the running spectral index confronting them with the cosmological observational data.
\subsection{Slow-roll parameters and inflationary observables}\label{subsec61}
The exploration of any inflationary scenario is essentially based on the values of related observables such as scalar spectral index of the curvature perturbations $n_s$, its  running $\alpha_s\equiv dn_s/d\ln k$, $k$ being  the absolute value of the wave number $\vec{k}$, the tensor spectral index $n_T$ and the tensor-to-tensor ratio $r$. The determination of the value of the observables detailed perturbation analysis and the we propose to scape from these complicated procedure by making use of the Einstein frame, where all the inflationary informations are driven by the so-called effective scalar potential $V(\phi)$. To do so its commode to define the slow-roll parameters $\epsilon$, $\eta$ and $\xi$ in terms of this potential and its derivatives as \cite{30deBOS,31deBOS}
\begin{eqnarray}
\epsilon\equiv\frac{M_p^2}{2}\left(\frac{1}{V}\frac{dV}{d\phi}\right)^2\;,\quad\quad \eta\equiv \frac{M_p^2}{V}\frac{d^2V}{d\phi^2}\;,\quad\quad
\xi^2\equiv \frac{M_p^4}{V^2}\frac{dV}{d\phi}\frac{d^3V}{d\phi^3}
\end{eqnarray} 
Also it has been shown in \cite{BOS} that the inflation ends when $\epsilon=1$. The approximated expressions for the observables are presented in \cite{31deBOS}
\begin{eqnarray}
r\approx 16\epsilon\;,\quad\quad n_s\approx 1-6\epsilon+2\eta\;, \quad\quad \alpha\approx 16\epsilon\eta-24\epsilon^2-2\xi^2\;,\quad\quad n_T\approx -2\epsilon
\end{eqnarray} 
Since we are dealing with a modified gravity, its obvious that the conformal transformation to the Einstein frame is not possible and one cannot define a scalar potential nor the potential slow-roll parameters. To do so, one can introduce the Hubble slow-roll parameters $\epsilon_n$ as
\begin{eqnarray}
\epsilon_{n+1}\equiv \frac{d\ln{\vert\epsilon_{n}\vert}}{dN},
\end{eqnarray}
where the initial value of this parameter is $\epsilon_0\equiv H_{ini}/H$ and the parameter $N$ known as the e-folding number is defined as $N\equiv \ln{\left(a/a_{ini}\right)}$, $a_{ini}$ being the scale factor at the beginning of the inflation and $H_{ini}$ the corresponding Hubble parameter. Hence one can express the first three $\epsilon$ as
\begin{eqnarray}
\epsilon_1=-\frac{\dot H}{H^2}\;,\quad\quad \epsilon_2=\frac{\ddot H}{H\dot H}-\frac{2\dot H}{H^2}\;, \epsilon_3=\left(H\ddot H-2\dot{H}^2\right)^{-1} \left[  \frac{H\dot H \dddot H-\ddot H \left(\dot{H}^2+ H\ddot H\right)}{H\dot H}-\frac{2\dot H}{H^2}\left(H\ddot H -2\dot{H}^2  \right)  \right],
\end{eqnarray}
such a way that the inflationary-related observables can now be written as
\begin{eqnarray}
r\approx 16\epsilon_1\;, \quad\quad  n_s\approx 1-2\epsilon_1-2\epsilon_2\;,\quad\quad \alpha_s\approx  -2\epsilon_1\epsilon_2-\epsilon-2\epsilon_3\;, \quad\quad n_T\approx-2\epsilon_1
\end{eqnarray}
According to (\ref{scaletq}) and (\ref{scaletausigma}), one gets
\begin{eqnarray}
\epsilon_1=\frac{1}{q}=\frac{1-3\sigma}{\sigma}\;,\quad\epsilon_2=0\;,\quad \epsilon_3=\frac{1}{q}=\frac{1-3\sigma}{\sigma}\;,
\end{eqnarray}
yielding
\begin{eqnarray}
r\approx \frac{16(1-3\sigma)}{\sigma}\;,\quad n_s\approx \frac{7\sigma-2}{\sigma}\;,\quad \alpha_s\approx 0\;,\quad n_T\approx \frac{2(3\sigma-1)}{\sigma}
\end{eqnarray}
\subsection{Obtaining inflationary unimodular $f(G)$ model according to the Planck results}\label{subsec62}
According to the Planck results \cite{2deBOS} one has $0.962\leq n_s\leq0.974$, leading to $52631\leq q\leq 76.923$. Thus, fixing $q=60$ that is $\sigma= 0.331$, one gets $n_s\approx 0.966$, $r\approx 0.266$, $\alpha_s=0$ and $n_T=0.033$, consistent with the Planck results. Now, still in agreement with the Planck results we will determine the constants appearing in (\ref{fdeG}) in order to obtain the $f(G)$ model that describes the inflation. We also note that according to the Planck results \cite{2debamba} one has $66.9 km/s/Mpc\leq H_0\leq 68.7 km/s/Mpc$ and fix it to $H_0=67.8 km/s/Mpc$ and the well known relation between the current values of the cosmic time  and the Hubble pointing out the age of the universe, i.e, $t_0\approx 2/3H_0$, one gets $t_0=9.832\times 10^{-3} Mpc.s/km$. Using the (\ref{scaletausigma}) and $A_0=1$, one gets $\tau_0=5.43\times 10^{-5} Mpc.s/km$. Using $\rho_0=3H_0^2$ and assuming that at this stage of inflation only the radiation is the dominated constituent of the universe (rigorously just after the inflation) i.e, $w=1/3$, one gets
\begin{eqnarray}
f(G)=1.186 \sqrt{G}-1.074\times 10^{-779}G^{47.285}-2.858\times 10^{205}G^{-11.571}+G
\end{eqnarray}
where we fix for simplicity $\mathcal{K}_1=\mathcal{K}_2=1$ and $\mathcal{K}_3=0$.

\section{Superbounce and loop quantum ekpyrotic cosmology from unimodular $f(G)$ gravity }\label{sec7}
In this section we propose to reconstruct the unimodular $f(G)$ model able to realize the superbounce and loop quantum ekpyrotic cosmology.
\subsection{Ekpyrotic scenario reconstruction from unimodular $f(G)$ gravity}\label{subsec71}
The task here is to find $f(G)$ model in the unimodular context which, in the large cosmic time limit, corresponds to the late-time era of the ekpyrotic scenario. We assume the well-known gravitational action as in \cite{86ekpyrotic,87ekpyrotic}
\begin{eqnarray}
\mathcal{S}=\int d^4x\sqrt{-g}\left[\frac{R}{2\kappa^2}+f(G)\right].\label{actionphi}
\end{eqnarray}
By making use of the auxiliary scalar field $\phi$ the previous action becomes
\begin{eqnarray}
S=\int d^4x\sqrt{-g}\left[\frac{R}{2\kappa^2}-V(\phi)-\xi(\phi)G\right]
\end{eqnarray}
The variation of the action (\ref{actionphi}) with respect to $\phi$ yields
\begin{eqnarray}
V'(\phi)+\xi'(\phi)G=0,
\end{eqnarray}
whose solution, if it exists, will be a functional dependence of $G$, i.e $\phi=\phi(G)$ and now substituting it into (\ref{actionphi}) lead to
\begin{eqnarray}
f(G)=-V(\phi(G))-\xi(\phi(G))G,\label{fGxiv}
\end{eqnarray}
and it is obvious that if we find $\phi(G)$, the algebraic $f(G)$ will be obtained straightforwardly.

\begin{eqnarray}
24A^{12}\mathcal{H}^3\xi'-24A^{12}\left(\mathcal{H}^2\mathcal{H}'+4\mathcal{H}^4\right)\xi+3A^6\mathcal{H}^2+V+\xi G=0\label{ekpyroticfrw1}\\
-8A^{12}\mathcal{H}^2\xi''-\left(16A^{12}\mathcal{H}\mathcal{H}'+88A^{12}\mathcal{H}^3\right)\xi'+24A^{12}\left(\mathcal{H}^2\mathcal{H}'+4\mathcal{H}^4\right)\xi-A^6\left(2\mathcal{H}'+9\mathcal{H}^2\right)-V-\xi G=0.\label{frw2}
\end{eqnarray}
By combining the above equations through a straight summation one gets the following equation

\begin{eqnarray}
-8A^{12}\mathcal{H}^2\xi''-16A^{12}\left(\mathcal{H}\mathcal{H}'+4\mathcal{H}^3\right)\xi'-2A^{6}\left(\mathcal{H}'+3\mathcal{H}^2\right)=0.\label{tobesolved2}
\end{eqnarray}
Since we are searching for ekpyrotic model we have to take large cosmic time limit of the scale factor assumed as $A(t)=A_0^{v/2}t^v$, corresponding, in the unimodular context, to
\begin{eqnarray}
A(t(\tau))=\bar{A}\tau^l,\quad l=\frac{v}{3v+1}\;,\quad \bar{A}=\left[(3v+1)\sqrt{A_0}\right]^l\label{baratau}
\end{eqnarray}

According to the previous expression of the scale factor in the unimodular context, the equation (\ref{tobesolved2}) takes the following form
\begin{eqnarray}
\xi''+\gamma_1(\tau)\xi'+\gamma_2(\tau)=0,
\end{eqnarray}
with
\begin{eqnarray}
 \gamma_1(\tau)=8\bar{A}l\tau^{l-1}+2(l-1)\tau^{-1}\;,\quad \mbox{and }\quad\gamma_2(\tau)=
\frac{3}{4}\bar{A}^{-6}\tau^{-6l}+\frac{1}{4l}\bar{A}^{-7}(l-1)\tau^{-7l}
\end{eqnarray}
From which, according to the large cosmic time, lead to the following expression of $\xi(\tau)$ and $V(\tau)$, as
\begin{eqnarray}
\xi(\tau)=\Gamma_1\tau^{2(1-3l)}\;, V(\tau)=\Gamma_2\tau^{-2(1-3l)} \label{xietv}
\end{eqnarray}
with
\begin{eqnarray}
\Gamma_1=\frac{(1-3l)^{6l-1}}{l(1-2l)\bar{A}^{3l}}\;,\quad \mbox{and}\quad \Gamma_2=24\frac{l^2(4l-1)(1-3l)^{2(3l-2)}}{1-2l}
\end{eqnarray}
Making use of (\ref{GBItau}) and (\ref{baratau}), one can express $\tau$in terms of $G$ as
\begin{eqnarray}
\tau=\Gamma_3G^{\frac{1}{4(3l-1)}}\;,\label{taudeG2}
\end{eqnarray}
where 
\begin{eqnarray}
\Gamma_3=\left[24\bar{A}l^3(4l-1)\right]^{\frac{1}{4(1-3l)}}
\end{eqnarray}
Now injecting the expression (\ref{taudeG2}) of $\tau$ into (\ref{xietv}), one obtains
\begin{eqnarray}
\xi(G)= \Gamma_1\Gamma_3^{2(1-3l)}G^{-1/2}    \;,\quad \mbox{and}\quad V(G)=\Gamma_2\Gamma_3^{2(3l-1)}G^{1/2},
\end{eqnarray}
such that, from (\ref{fGxiv}),  the algebraic $f(G)$ ekpyrotic model reads
\begin{eqnarray}
f(G)=-\Gamma_1\Gamma_3^{2(1-3l)}G^{-1/2}-\Gamma_2\Gamma_3^{2(3l-1)}G^{3/2}
\end{eqnarray}
\subsection{Superbounce Reconstruction from unimodular $f(G)$ Gravity}\label{subsec72}
In this case the scale factor is given by \cite{11deekpyrotic}
\begin{eqnarray}
A(t)\propto (t_*-t)^{2/c^2}
\end{eqnarray}
where $t_*$ denotes the big crunch time, and $c$ a parameter constrained to $c>\sqrt{6}$ \cite{11deekpyrotic}. In the context of unimodular gravity, one gets
\begin{eqnarray}
A(t(\tau))\propto (\tau_*-\tau)^{\frac{2}{6+c^2}}\label{Adetaubounce}
\end{eqnarray}
In this case, for simplicity and for large cosmic time $t$, i.e near the bounce, the $\tau$ dependent functions $\xi(\tau)$ and $V(\tau)$ behave as
\begin{eqnarray}
\xi(\tau)\propto (\tau_*-\tau)^{\frac{2c^2}{6+c^2}}\;,\quad V(\tau)= (\tau_*-\tau)^{\frac{-2}{6+c^2}}
\end{eqnarray}
On the other hand, from (\ref{Adetaubounce}) and (\ref{GBItau}), one expresses $\tau$ in terms of $G$ as
\begin{eqnarray}
\tau(G)=G^{-\frac{6+c^2}{4c^2}},
\end{eqnarray}
such that the algebraic $f(G)$ model near the bounce reads
\begin{eqnarray}
f(G)=\mathcal{Z}_1G^{-1/2}+\mathcal{Z}_1G^{3/2}
\end{eqnarray}
where $\mathcal{Z}_1$ and $\mathcal{Z}_2$ are integration constants depending on the parameter $c$.

\section{Conclusion}\label{sec8}
In this paper we explored a type of modified GR theory, namely $f(G)$ theory of gravity in the unimodular context, where $G$ denotes the GB invariant. In such a theory an important condition is applied to the metric tensor constraining its determinant to a number or a specific function. Here we work fixing it to $1$ and considered both appropriated BI and FRW universes. In a first step the task is to reconstruct the unimodular $f(G)$ models considering de Sitter and power-law solutions for the scale factors. The resulting models are different from the standard $f(G)$ model (without considering unimodular formalism). In the second step we focus our attention to the model provided by the power-law solutions in the FRW universe and searched for the input constants that can constrain it to an inflationary model. To do so, we proceed to the determination of the slow-roll parameters and the corresponding observational indices from the unimodular $f(G)$ field equations, that appear as function of the input parameters. According to the recent Planck  results we calculate accordingly the input parameters, obtaining an unimodular $f(G)$ inflationary model.\par
On the other hand we investigate, still in the unimodular  $f(G$ context, the superbounce and the loop quantum cosmology ekpyrotic paradigms. As it is well known, bouncing solutions can be an alternative to inflation and thus it appears interesting to search for  related $f(G)$ model. To do so, we considered well known standard cosmic time depending scale factors able to lead to the bouncing cosmology in a case and the ekpyrotic cosmology in other. The correspondent scale factors in the unimodular context have been determined, namely in term of the auxiliary time $\tau$. Through the unimodular formalis and the related field equations, the bouncing and ekpyrotic $f(G)$ models have been reconstructed. 

\vspace{0.25cm}
{\bf Acknowledgement:} The author thanks Profs. S. D. Odintsov,  V. Oikonomou and A. Yu. Petrov for useful comments an
suggestions.

\end{document}